\documentclass[fleqn,usenatbib]{mnras}

\usepackage{newtxtext,newtxmath}

\usepackage[T1]{fontenc}
\usepackage{ae,aecompl}

\usepackage{lscape}

\usepackage{graphicx}	
\usepackage{amsmath}	
\usepackage{amssymb}	

\def\Swift{{\em Swift}}
\newcommand\xmm{{\em XMM-Newton}}
\newcommand\ergscm{$\rm erg\,cm^{-2}\,s^{-1}$}

\newcommand\src{SWIFT\,J0658.0-1746}

\title[The Polar 2PBC\,J0658.0-1746]{2PBC\,J0658.0-1746: a hard X-ray eclipsing polar in the orbital period gap}

\author[F. Bernardini et al.]
{F.~Bernardini,$^{1,2,3}$\thanks{E-mail:federico.bernardini@inaf.it} 
D.~de Martino,$^{2}$ 
K.~Mukai,$^{4,5}$
M.~Falanga,$^{6,7}$
N.~Masetti,$^{8,9}$
\\
$^1$ INAF - Osservatorio Astronomico di Roma, via Frascati 33, I-00040 Monteporzio Catone, Roma, Italy \\
$^2$ INAF $-$ Osservatorio Astronomico di Capodimonte, Salita Moiariello 16, I-80131 Napoli, Italy\\
$^3$ New York University Abu Dhabi, Saadiyat Island, Abu Dhabi, 129188, United Arab Emirates\\
$^4$ CRESST and X-Ray Astrophysics Laboratory, NASA Goddard Space Flight Center, Greenbelt, MD 20771, USA\\
$^5$ Department of Physics, University of Maryland, Baltimore County, 1000 Hilltop Circle, Baltimore, MD 21250, USA\\
$^6$ International Space Science Institute (ISSI), Hallerstrasse 6, 3012 Bern, Switzerland\\
$^7$ International Space Science Institute Beijing, No.1 Nanertiao, Zhongguancun, Haidian District, 100190 Beijing, China\\
$^8$ INAF - Osservatorio di Astrofisica e Scienza dello Spazio, Via Gobetti 93/3, I-40129, Bologna, Italy\\
$^9$ Departamento de Ciencias F\'isicas, Universidad Andr\'es Bello, 
Fern\'andez Concha 700, Las Condes, Santiago, Chile}

\date{Accepted XXX. Received YYY; in original form ZZZ}

\pubyear{2018}

\begin{document}
\label{firstpage}
\pagerange{\pageref{firstpage}--\pageref{lastpage}}
\maketitle

\begin{abstract}
The hard X-ray source 2PBC\,J0658.0-1746 was  proposed as an eclipsing 
magnetic cataclysmic variable of the polar type, based on optical follow-ups. 
We present the first spectral and timing analysis at X-ray energies with 
XMM-Newton, complemented with archival X-ray, optical, IR photometry and 
spectroscopy. The X-ray emission shows bright and faint phases and total 
eclipses recurring every 2.38\,h, consistent with optical properties. This 
firmly identifies 2PBC\,J0658.0-1746  as an eclipsing polar, the second hard 
X-ray selected in the orbital period gap.  The X-ray orbital modulation 
changes from cycle-to-cycle and the X-ray flux is strongly variable over 
the years, implying a non-stationary mass accretion rate both on short and 
long timescales. The X-ray eclipses allow to refine the orbital ephemeris 
with period 0.09913398(4)\,d, and to constrain the binary inclination 
$79^{\rm\,o}\lesssim i \lesssim 90^{\rm\,o}$ and the mass ratio 
0.18$\rm <M_2/M_{WD}<$0.40. A companion mass M$_{2}=0.2-0.25\rm\,M_{\odot}$ 
with a radius R$_{2}=0.24-0.26\rm\,R_{\odot}$ and spectral type $\sim$M4, 
at D$=209^{+3}_{-2}\rm\,pc$, is derived. A lower limit to the white dwarf 
mass of $\sim0.6\,\rm\,M_{\odot}$ is obtained from the X-ray spectrum. 
An upper limit to the magnetic colatitude, $\beta \lesssim 50^{\rm\,o}$, 
and a shift in azimuth, $\psi\sim14^{\rm\,o}$, of the main accreting 
pole are also estimated. The optical/IR spectral energy distribution 
shows large excess in the mid-IR due to lower harmonics of cyclotron 
emission. A high-state mass accretion rate 
$\rm\,\sim0.4-1\times10^{-10}\,M_{\odot}\,yr^{-1}$, lower than that 
of cataclysmic variables above the gap and close to that of systems 
below it, is estimated. With 2PBC\,J0658.0-1746, the number of hard 
X-ray selected polars increases to thirteen members, suggesting that 
they are not as rare as previously believed.

\end{abstract}

\begin{keywords}
Novae, cataclysmic variables - white dwarfs - X-rays: individual: 
SWIFT\,J0658.0-1746 (aka 2PBC\,J0658.0-1746, 1RXS\,J065806.3-1744, 2MASS\,06580591-1744249, WISE\,J065805.87-174424.5 )
\end{keywords}



\section{Introduction}

AM Herculis stars or polars, are magnetic cataclysmic variables (mCVs), short orbital period (up to a few hours) binary systems, hosting a strongly magnetic (10--230 MG) white dwarf (WD) primary that accretes matter via Roche lobe overflow from a main sequence late-type secondary \citep[see][]{Cropper90,ferrario15}. The magnetic field is strong enough to synchronise or quasi-synchronise the WD rotation at the binary orbital period ($\rm P_{orb}= P_{\Omega}$)
so that material lost from the companion directly proceeds in a stream-like flow  along the field lines from the inner Lagrangian point (L1) all the way down to the polar magnetic cap(s) on the WD surface,  preventing the formation of an accretion disk. In the magnetically confined accretion flow, matter reaches supersonic velocities and a stand-off shock forms close to the WD surface \citep{aizu73}. The post shock region (PSR) is hot (kT$\sim10-50$ keV) and matter below cools down emitting thermal Bremsstrahlung (hard X-rays), and cyclotron radiation, emerging in the  optical/near-infrared (nIR) band \citep{wu94,cropper99}. The efficiency of the two mechanisms depends on the magnetic field strength and local mass accretion rate, with the higher magnetic field systems being cyclotron dominated. 
Due to reprocessing of the hard X-rays and cyclotron radiation at the WD surface or heating from blobby accretion \citep{konig06}, an optically thick (blackbody-like) soft X-ray component can be observed. This was believed to be almost ubiquitous in polars during the $ROSAT$ era.  However, more recent observations with \xmm\ showed an increasing number of such sources without a distinct soft X-ray component \citep[see e.g.][]{ramsayandcrop04,bernardini14,bernardini17}. 

The X-ray (and frequently also optical) light curves of polars show periodic variability at the WD spin/orbital period, a clear signature of magnetic accretion. Their shape is extremely peculiar, usually showing a double peak with a broad, V-shaped dip due to partial obscuration of the emitting spot by the accretion stream \citep[e.g.][]{ramsay04,bernardini14}. On top of short-term (hours) periodic variations, polars also show long-term (months--years) variability with high and low accretion states. Magnetic spots on the donor surface could temporarily fill the Lagrangian point \citep[L1,][]{livio94} producing variable mass-transfer rates from the secondary \citep[e.g.][]{hessman00}.

Polars represent the major subclass of mCVs, with $\sim130$ systems identified so far \citep[][]{ritter_kolb,ferrario15}\footnote{For \citep{ritter_kolb}, see catalog version 7.24, more details at  https://wwwmpa.mpa-garching.mpg.de/RKcat/.}, mainly discovered in the soft X-rays through the $ROSAT$ \citep{Schwope02} and \xmm\ \citep[e.g.][]{ramsay09} serendipity surveys and in optical photometric surveys such as the Sloan Digital Sky Survey \citep[SDSS; e.g][]{schmidt08} or the Catalina Real-time Transient Survey \citep[CRTS; e.g.][]{breedt14}. The lower-field ($\rm B<10-20$ MG) asynchronous systems, the so-called intermediate polars, amount to about 70 systems. Thus, mCVs represent about 20--25$\%$ of the whole CV class. The high incidence of magnetism among WDs in CVs, compared to $\sim6-10\%$ of isolated magnetic WDs \citep{ferrario15}, would either imply CV formation is favoured by magnetism or CV production enhances magnetism \citep{tout08}.
Moreover, to explain their lower mass accretion rates with respect to non-magnetic CVs, it was proposed that mCVs, especially the polars, would suffer of reduced efficiency of the magnetic braking mechanism, operating as angular momentum loss (AML) above the 2--3\,h orbital period gap, due to the strong coupling of the WD and companion magnetic fields \citep{wickramasinghe_wu94,wickramasinghe00,ferrario15}.

Only recently, thanks to the hard X-ray surveys conducted by $INTEGRAL$ with IBIS-ISGRI \citep{bird16} and by the Neil Gehrels \Swift\ observatory (hereafter \Swift) with BAT \citep{oh18}, the number of discovered mCVs has substantially increased, representing $\sim25\%$ of galactic hard X-ray sources. While the majority are found to be of the intermediate polar-type, likely due to the dominance of Bremsstrahlung as main cooling mechanism, the number of polars is slowly, but steadily increasing amounting (excluding this work) to twelve systems \citep{bernardini14,mukai17,bernardini17}, indicating that hard X-ray polars are not so rare as previously thought. Whether IPs only or also the polars are contributors to the low-luminosity population of Galactic X-ray sources is still under debate \citep[][]{reis13,pretorius14}.   

Here is reported, in the framework of an identification programme with the \xmm\ satellite of hard X-ray selected CV candidates \citep[see e.g.][and references therein]{bernardini12,bernardini13,bernardini17,bernardini19}, the first X-ray study of \src\ (hereafter J0658), discovered as an unidentified source in the \Swift/BAT survey \citep{cusumano14,oh18}. It was proposed as a mCV because of its optical spectral characteristics \citep{rojas17}, until follow-up optical photometery in 2017-2018 revealed that it is an eclipsing system with a period of 0.0991370(3) days and large orbital modulation characteristic of polars \citep[][]{halpern18}. The \xmm\ observation, carried out a few months later in 2018, is here reported together with archival \Swift/XRT and \Swift/BAT light curves and spectra. These confirm the 2.38\,h orbital period and that J0658 is a hard X-ray eclipsing polar. It is thus the $\rm 13^{th}$ hard X-ray discovered so far and the $\rm 2^{nd}$ falling in the middle of the 2--3\,h orbital period gap, together with SWIFT\,J2218.4+1925 \citep[][and reference therein]{bernardini14}. 

In Section \ref{sec:data} we report the observations and data reduction procedures, in Section \ref{sec:results} the timing and spectral analysis, in Section \ref{sec:discus} we discuss the physical properties of this system in the context of mCVs, and in Section \ref{sec:conc} we highlight our main conclusions. 

\section{Observation and data reduction}
\label{sec:data}

\subsection{\textit{XMM-Newton} observations}
\label{sec:xmmdata}

J0658 has been observed for $\sim33$ ks on 2018 September 19 by \xmm\ with 
the European Photo Imaging Cameras \citep[EPIC: PN, MOS1 and MOS2][]{struder01,turner01,denherder01} 
as main instruments, complemented with simultaneous Optical Monitor  \citep[OM;][]{mason01} V-band photometry and RGS spectroscopy \citep{struder01}. The latter data, due to the faintness of the source, were of poor S/N for a useful analysis.
The details of the observation are reported in Table \ref{tab:observ}. Data were processed using the 
Science Analysis Software (\textsc{SAS}) version 17.0.0 and the latest calibration files available in 2019 May. 

Source photon event lists and spectra for EPIC cameras were extracted from a circular region of radius 30 arcsec. 
The background was extracted in the same CCD where the target lies. The observation did not contain epochs dominated by particle background, so the whole exposure was used for the analysis.

Background-subtracted PN and MOSs light curves were produced with the task \textsc{epiclccorr} in the whole 0.3--12 keV energy band, and in several 
energy sub-bands. The event arrival times were corrected to the Solar System barycentre by using the 
task \textsc{barycen}. Before fitting, spectra were rebinned using \textsc{specgroup}. A minimum of 
30 and 25 counts in each bin for PN and MOSs, respectively, and a maximum oversampling of the energy 
resolution by a factor of three were set. Three time resolved spectra were also extracted, corresponding to the two highest bright phases, two less intense bright phases, and all faint phases of the spin/orbital cycle present in the exposure, respectively (see Section \ref{sec:timing} and \ref{sec:spec}). 
The response matrix and the ancillary files were generated using the tasks 
\textsc{rmfgen} and \textsc{arfgen}, respectively. PN and MOSs spectra were fitted together by using \textsc{Xspec} version 12.10.1f package \citep{arnaud96}.

The OM was operated in fast window mode using the
V-band (5100--5800 \AA) filter simultaneously to the EPIC cameras. 10 exposure of $\sim2.9$ ks each were performed. The background subtracted light curve was generated with the 
task \textsc{omfchain} with a bin time of 100\,s and then the Solar System barycentric correction was applied. 

\subsection{The \Swift\ observations}

J0658 was observed twice by \Swift/XRT \citep{burrows05} a first time in 2009 February (1.1 ks) and a second time in 2013 June (1.6 ks). XRT (0.3--10 keV) light curve and spectra were also extracted using the products generator available at Leicester \Swift\ Science Centre \citep{Evans09}. Since the source flux and spectral shape are consistent during the two epochs, we produced and analysed a single averaged spectrum.

\Swift/BAT \citep{barthelmy05} has built up an all-sky map of the hard X-ray sky (14--195 keV), thanks to its wide field of view. The eight-channel spectra (14--195 keV), response file, and light curve from the first 105 months of BAT monitoring \citep{oh18} were downloaded from the publicly available archive at the NASA's Goddard Space Flight Center website\footnote{https://swift.gsfc.nasa.gov/results/bs105mon/}. We restricted the spectral analysis to E$<$80 keV.


\section{Data analysis and results}
\label{sec:results}

\begin{table*}
\caption{Summary of main observation parameters for all instruments. Uncertainties are at $1\sigma$ confidence level.}
\begin{center}
\begin{tabular}{cccccccc}
\hline 
Source Name            & Telescope            & OBSID        & Instrument    & Date        & UT$_{\rm start}$ & T$_{\rm exp}$ $^a$ & Net Source Count Rate\\
Coordinates (J2000)$^{b}$&               &              &        & yyyy-mm-dd      & hh:mm & ks      &    c/s                  \\
\hline
\src\    & \xmm\   &  0820330701          & EPIC-PN$^c$   & 2018-09-19  & 18:29 & 32.9 & $0.093\pm0.002$ \\
         &         &                     & EPIC-MOS1$^c$ & 2018-09-19  & 18:21 & 33.3 & $0.023\pm0.001$     \\ 
RA=06:58:05.873	   &                      &              & EPIC-MOS2$^c$ & 2018-09-19  & 18:21 & 33.3 & $0.025\pm0.001$     \\               
Dec=-17:44:24.40    &                      &              & OM-V$^d$      & 2018-09-19  & 18:26 & 29.0 & $18.5\pm0.2^e$    \\
         & \Swift\  &  038850001            & XRT       & 2009-02-13  & 05:30	      & 1.1 & $0.24\pm0.01$ \\ 
          & \Swift\ &  047443001            & XRT$^f$   & 2013-06-07   &	02:27         & 1.6 & $0.21\pm0.01$ \\ 
                                      & \Swift\              &              & BAT$^g$       &             &	      & 185500 & $1.4\pm0.3\times 10^{-5}$ \\ 
\hline              
\end{tabular}
\label{tab:observ}
\end{center}
\begin{flushleft}
$^a$ Net exposure time. The PN ontime is 23 ks only.\\
$^b$ Coordinates of the optical counterpart. \\
$^c$ Small window mode (medium filter applied for MOS1 and MOS2, thin filter applied for PN). \\
$^d$ Fast image mode. The central wavelength of the V filter is 5430 \AA. \\ 
$^e$ OM V magnitude in the Vega system. \\
$^f$ This pointing consists of two PC mode snapshots of 0.8 ks each, the second starting on 2013-06-08 at 08:55. \\
$^g$ All available pointings collected from 2004 December to 2013 September are summed together. \\
\end{flushleft}
\end{table*}

\subsection{Timing analysis} 
\label{sec:timing}

First the background subtracted \Swift/BAT lightcurve of J0658 was inspected and found to show clear long-term (years) variability, with a broad ($FWHM\sim5$ months) peak centered at t$\sim90$ months (2012 May). The 14--195 keV flux at the peak is $\sim7\times10^{-11}$ \ergscm, while the average flux over $\sim7$ yrs outside the peak is $\sim4.5\times10^{-12}$ \ergscm (factor $\sim16$ lower). The Swift/XRT, 0.3--10 keV, flux level in 2009/2013 was $1.8\times10^{-11}$ \ergscm assuming a simple power-law spectral model (Section \ref{sec:spec}).

The average magnitude during the OM pointing is V$=18.5\pm0.2$ mag, 
which is $\sim2.2$ mag fainter than  what obtained interpolating the 
B and R magnitudes \citep{halpern18} from the USNO B1.0 catalog 
\citep{monet03}, that were taken over several decades. 
J0658 is found 1.3\,mag fainter with respect to the USNO B1.0 catalog also in the V-band (17.2\,mag), when using the {\it Gaia} DR2 G-band magnitude and red (Rp) and blue (Bp) colors \citep{gaia18a}\footnote{The {\it Gaia} DR2 data covers a 22 months period, starting from July 2014.} along with the color-transformation by \cite{evans18}. Furthermore, when later observed in December 
2017 \citep{halpern18} the R-band photometry was consistent with the R-band magnitude obtained from the {\it Gaia} color-transformation (R$\sim$16.7).
J0658 WAS also detected in the Sloan Digital Sky Survey (SDSS) DR13 \citep{blanton17} and observed in 2008 at g=17.2, similarly to the {\it Gaia} level. All this indicates that J0658 faded after December-February 2018 and is highly variable on long timescales.

\begin{figure}
\begin{center}
\includegraphics[angle=-90,width=9cm]{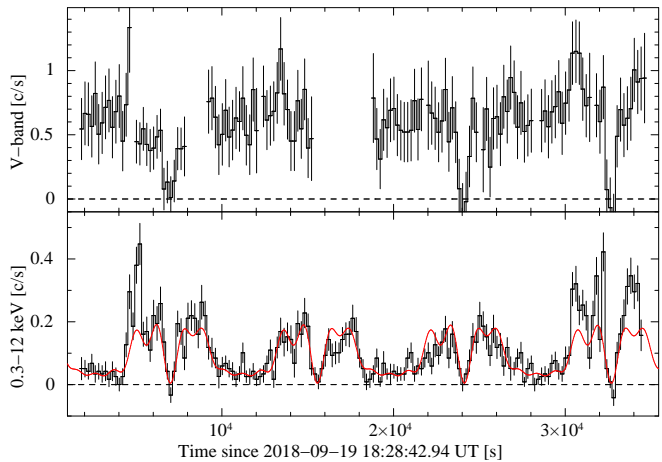} 
\caption{2018 September 19 OM V-band (upper panel) and PN 0.3--12 keV (lower panel) background subtracted light curves binned at 200 s, where for plotting purposes is also shown the result of a fit made with 7 sinusoids, the orbital period (8565 s) and its first six harmonics (red curve).}
\label{fig:lc-xmm}
\end{center}
\end{figure}

The $0.3-12$ keV background subtracted PN light curve (Figure \ref{fig:lc-xmm}) 
shows a periodic modulation with a structured bright phase and a faint phase which are typical of polars (four cycles are covered). The bright phase is characterised by a total eclipse that separates two unequal peaks. The modulation changes in shape from cycle to cycle, with the first and last bright phases (hereafter B) brighter than the other two (b). The count rate during the faint phases (f) is above zero. The power spectrum, computed in the $0.3-12$ keV range, shows a strong peak at $\sim10^{-4}$ Hz together with several harmonics up to the $\rm 7{th}$. A first estimate of the period of the fundamental harmonic is obtained by simply fitting the EPIC averaged (PN, MOS1, and MOS2) $0.3-12$ keV light curve with a series of sinusoids consisting of the fundamental 
plus the first 6 harmonics and gives $P=8565\pm34$ s. This is perfectly consistent with that derived from optical photometry \citep[][]{halpern18}.
All uncertainties are hereafter at $1\sigma$ confidence level, if not otherwise specified.

The V-band OM light curve does not show clear signs of bright and faint phases and so a detectable orbital modulation, but the eclipses (only three covered) are clearly observed (Figure \ref{fig:lc-xmm}), where $V\sim 19.4\pm$0.4\,mag. \cite{halpern18} found instead that in 2017 December, the light curve of J0658 shows a periodic modulation with a
double-peaked maximum at R$\sim$15.5\,mag rising from a baseline of R$\sim$17.3 mag. All this indicates that in 2018 September J0658 was in a low-intermediate state. 

\subsubsection{The eclipse}
\label{subsubsec:ephem}
\begin{figure}
\begin{center}
\includegraphics[angle=0,width=7cm]{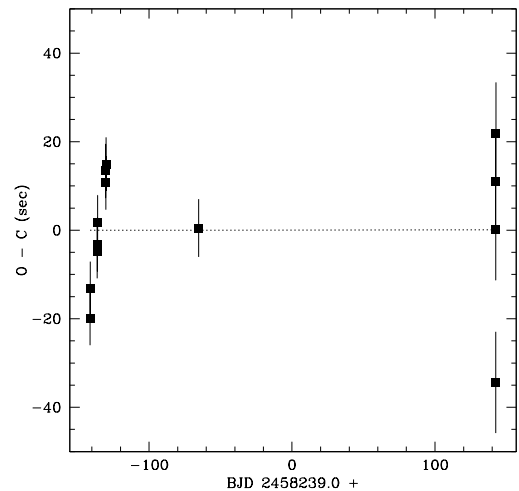} 
\caption{Deviations of the times of eclipse egress from the linear ephemeris (dotted line) obtained by fitting the nine optical and the four X-ray egresses (Section \ref{subsubsec:ephem}).}
\label{fig:O_C}
\end{center}
\end{figure}

The coverage of four eclipses in the X-ray band 
allows us to derive a new, improved, orbital ephemeris. The times of eclipse centres were measured using a Gaussian fit and cross-checked by manual inspection on the 50s-binned background subtracted PN light curve. The ingress(egress)  times were measured
from two points that have a flux that is half of the average value immediately before(after) the eclipse. The X-ray eclipse has 100$\%$ depth (count rates going to zero) and all four measures have consistent length with an average value of 10.73$\pm$0.43\,min. The total length of the eclipse (time of first to the fourth/last contact)
is  13.89$\pm$0.06\,min for the first and third eclipse and slightly longer, 14.8$\pm$0.1min, in the second and last one.  The OM V-band eclipses, as measured on the 100s-binned light curve, last 9.64$\pm$0.48\,min (second to third contact) and 13.6$\pm$0.6\,min (first to last contact). The X-ray and optical eclipses lengths are consistent within their uncertainties and with the previous optical observations by \cite{halpern18}. Instead, the drop in V magnitude  with respect to the average out-of-eclipse level, is $\sim$1\,mag, much smaller than previously observed.
In order to obtain an improved ephemeris, the nine eclipse egresses reported by 
\cite{halpern18}\footnote{We did not use the times of ingress since only eight are reported by 
\cite{halpern18}} and the four times of X-ray eclipse egress were used. 
 The OM V-band eclipses are not used since the light curve has a lower S/N. A linear regression using the thirteen measures gives: $\rm T^{egress}$(BJD)=2458381.333958(65) and $\rm P_\Omega$=0.09913398(4)\,d. The observed-minus-calculated residuals were inspected against trends finding that a constant period
gives an acceptable fit. The excursions on average are within 20\,s around zero, with only one within 34\,s (Figure \ref{fig:O_C}). 

\begin{figure*}
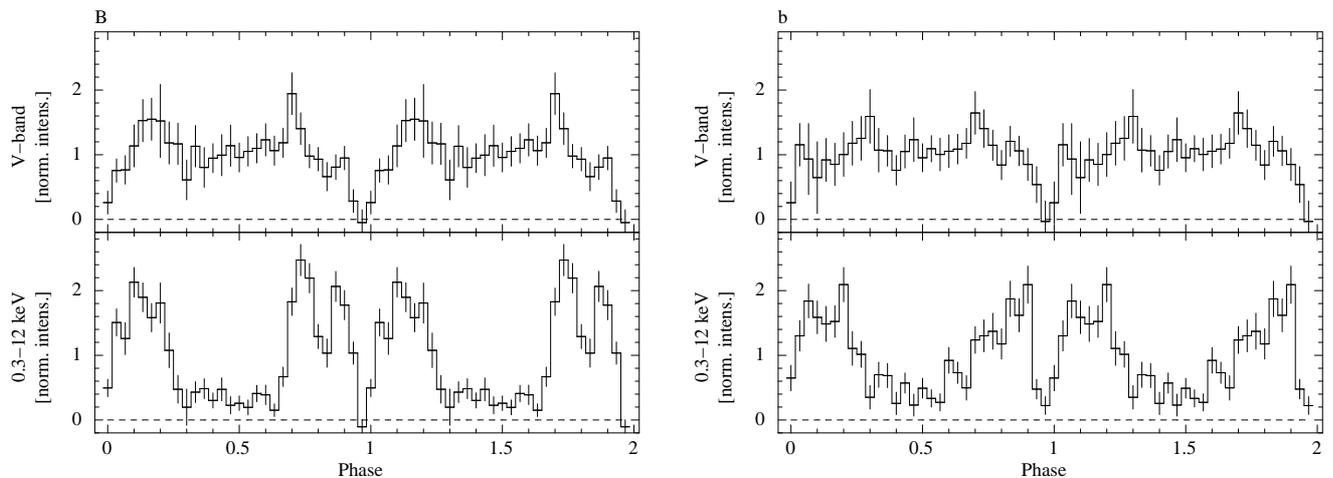

\begin{center}
\begin{tabular}{cc}
\includegraphics[angle=270,width=3.5in]{pulse_Max_bright.eps} 
\includegraphics[angle=270,width=3.5in]{pulse_max_faint.eps} 
\end{tabular}
\caption{Left: Background-subtracted V-band (top) and PN 0.3--12 keV (bottom) light curves, folded using the improved ephemeris (Table \ref{tab:binary})
for the interval covering the two stronger bright phases (B). 
Two cycles are shown for plotting purposes. Right: The same as left panel, but for the interval covering the two less intense bright phases (b). Note the presence of a dip before the X-ray eclipse, only during B, at $\phi\sim0.82$.}
\label{fig:pulse}
\end{center}
\end{figure*}


The background-subtracted (PN, MOS1, and MOS2) light curves in the 0.3--12\,keV and V-band were then folded with the improved ephemeris. Since the source shows cycle-to-cycle variability the folding was performed on two intervals encompassing the two strongest bright phases (B) and the two less intense (b) \footnote{More specifically interval B extends between MJD 58380.771 and 58380.875 and between MJD 58381.079 and 58381.550, whilst interval b extends between MJD 58380.884 and 58381.088.}. In Figure \ref{fig:pulse} the X-ray and V-band folded orbital light curves are shown. Notably is the presence in the X-ray orbital modulation during B of a dip preceding the eclipse (at $\phi_\Omega\sim0.82$), not seen during b. The pulsed fraction (PF)\footnote{The pulsed fraction is here defined as the semi-amplitude of the modulation divided by its average value.} of  the fundamental harmonic (excluding the eclipse interval) in the X-rays is $57\pm7\%$ during B. In the optical the PF is consistent with zero, and only the eclipse is clearly present.
The second, b,  interval  is characterised by a bright phase with no clear presence of a dip in both bands. The orbital modulation in the X-ray band PF$=54\pm7\%$ and consistent with zero in the optical band. Inspection of hardness ratios folded at the orbital period in the two intervals do not show any variability, neither at the dip feature, likely because of the low S/N. These findings could indicate that the source has a highly variable accretion rate on timescales of hours and when brighter the accretion stream passing through the line-of-sight of the observer partially obscures the accreting region. 

\subsection{Spectral analysis}
\label{sec:spec}

The broad-band (0.3--80 keV), time-averaged, \xmm\ EPIC plus BAT combined, spectrum was first fitted using an optically thin plasma (\textsc{cemek} in \textsc{Xspec}), a model that accounts for the run of emission measure with temperature in PSR of mCVs, \citep[see e.g.][]{mukai17,bernardini18}. While this model well fits the data, the inter-calibration constant, used to account for instrument calibration discrepancies and spectral variability, is extremely large, 60 for the BAT spectrum. This indicates that J0658, when observed by \xmm, is much fainter than the BAT average. This is consistent with the observed low optical level of the source. Moreover, the combined spectra do not provide substantial improvement in the fit and particularly the maximum temperature of the \textsc{cemek} model.  We then here report the results from the fits using  the 0.3--10 keV spectra of the three EPIC cameras. 

\begin{figure*}
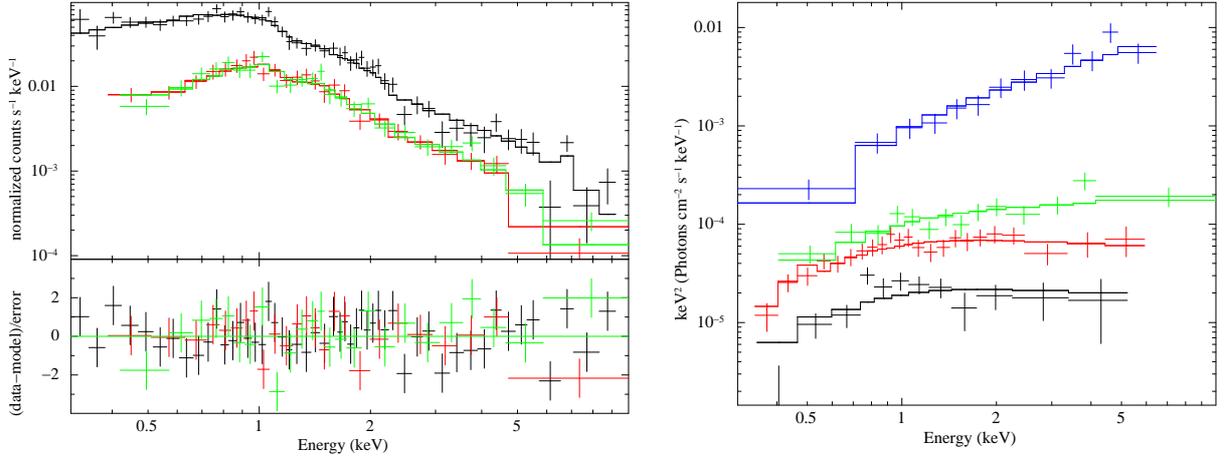

\begin{center}
\begin{tabular}{cc}
\includegraphics[angle=270,width=3.4in]{spec_xmm.eps} 
\includegraphics[angle=270,width=3.0in]{xrt_xmm.eps} 
\end{tabular}
\caption{Left: 0.3--10 keV count, averaged, spectrum fitted using \textsc{cemek} (Table \ref{tab:spec}). PN spectrum is in black, MOS1 in red, MOS2 in green. Residuals are shown in the lower panel. Right: Unfolded 0.3--10 keV spectra using a simple, phenomenological, power-law model. XRT spectrum is in blue, PN interval B in green, PN interval b in red, PN interval f in black. The spectra are plotted together to highlight the source flux variability, and spectral hardening, with increasing X-ray flux.}
\label{fig:spec}
\end{center}
\end{figure*}

We linked all models parameters among different spectra, leaving free to vary only the cross-normalization constant (fixed to 1 for the PN). We obtained statistically acceptable fits (Table \ref{tab:spec}, and Figure \ref{fig:spec}) using both \textsc{cemek} ($\chi_{\nu}^{2}=1.04$, 94 d.o.f) and a model consisting of two \textsc{mekals}, absorbed only by a total covering column \textsc{tbab} ($\chi_{\nu}^{2}=1.03$, 93 d.o.f.). The absorber accounts for the absorption from the Galactic interstellar medium. Metal abundances ($\rm A_Z$) with respect to Solar were set to that of \citep{wilms00} and left free to vary. Due to the source faintness (and low S/N), the fluorescent 6.4 keV Fe emission line, usually observed in mCVs, is not detected in J0658. Moreover, neither a partial covering absorbing column, usually present in these accreting systems due to self-obscuration from the accretion stream, is needed to fit the data. This in turn explains why the PF does not change with the energy interval. 
The emission measure in \textsc{cemek} follows a power-law in temperature $dEM=(T/T_{max})^{a-1}\,dT/T_{max}$, where T$_{max}$ can be considered as a lower limit to the real shock temperature. The best fit gives kT$_{max}=19\pm_{5}^{10}$ keV and $\alpha=0.56\pm^{0.19}_{0.23}$. The model with two mekals (cold and hot, respectively) gives instead kT$_{c}=0.66\pm0.04$ keV and kT$_{h}=6.1\pm_{0.6}^{0.9}$ keV. The total absorber column density is a factor of about 10 lower than the Galactic value in the direction of the source, 3.4--3.7$\times10^{21}$\,cm$^{-2}$ \citep[][]{dickey90,kalberla05}, consistent with a close-by source (see Sect \ref{sec:discus}). The fits also do not require a black-body component, adding this source to the increasing number of polars without a distinct soft component \footnote{Adding a blackbody with kT$=0.05$ keV (fixed), we derive a $3\sigma$ upper limit to the 0.3--10 keV flux, which is $<8\times10^{-14}$ \ergscm.}. 

The 0.3--10 keV spectra of interval B, b, and f, were also analysed. The spectra of the three EPIC cameras, for each interval separately, were fitted simultaneously using the best-fitting average spectral models by linking all the model components with the exception of N$_{\rm H_{Tbab}}$ and A$_{\rm Z}$ that were set to their average spectral best fitting values. 
For what concerns the \textsc{cemek} model, first the spectrum of interval B was fitted with both kT$_{max}$ and $\alpha$ left free to vary, but they resulted poorly constrained: kT$_{max}=22.6\pm_{8}^{24}$ keV,  $\alpha=0.63\pm0.20$, and n$=6\pm1\times10^{-4}$ ($\chi^{2}_{\nu}$=1.08, 54 d.o.f.).
Then,  $\alpha$ was set to its average spectrum best-fitting value. The fits show that kT$_{max}$ increases (the source gets harder) from a minimum of $10.3\pm_{2.5}^{3.7}$ keV to a maximum of $27.4\pm_{4.3}^{5.4}$ keV (the same does the normalization of \textsc{cemek}) as the X-ray flux increases (Table \ref{tab:spec_vs_t}). 
In the case of the model made with two \textsc{mekals}, instead, the temperature of the two optically thin plasma remain constant within uncertainty, but their normalizations increase with the flux. In particular that of the hot component significantly increases from a minimum of $0.5\pm0.1\times10^{-4}$ to a maximum of $3.3\pm0.2\times10^{-4}$.

Given the source long-term X-ray variability, the XRT spectrum 
averaged over the pointings in 2009 and 2013 was also analysed. First the broad-band spectrum, including BAT, was fitted using the same models in Table \ref{tab:spec}. The inter-calibration constant for BAT is $\sim0.20$, implying that J0658 when observed by XRT is brighter than the BAT average. Moreover, kT$_{max}$ is unconstrained, reaching the model maximum value (100 keV) \footnote{Using \textsc{pwab} (normally adopted when CV spectra look unusually hard using other simpler absorption model), instead of \textsc{tbab}, kT$_{max}$ remains unconstrained.}. This also suggests that J0658 spectrum is very hard when observed by XRT. The XRT spectrum was then fitted alone, first using the same models in Table \ref{tab:spec}, and kT$_{max}$ (or kT$_{h}$) is found to be unconstrained\footnote{The same is true if \textsc{pwab} is used.}. Then, a simple power-law model was used and gives: N$_{h}=0.05$ (fixed), $\Gamma=0.9\pm0.1$, norm$=1.0\pm0.1\times10^{-3}$, F$_{\rm 0.3-10\, keV}=1.8\pm0.2\times10^{-11}$ ($\chi_{\nu}^{2}=0.62$ 11 d.o.f..). This shows that J0658, when observed by XRT, is a factor of $\sim60$ brighter than in 2018. It also shows that it is much harder ($<\Gamma_{2018}>=1.74\pm0.04$), confirming the hardening trend with increasing X-ray fluxes (Figure \ref{fig:spec}).

\begin{table*}
\caption{Best fit models to the average 0.3--10 keV spectrum. Uncertainties are at $1\sigma$ confidence level.
The unabsorbed 0.3--10 keV fluxes are also reported.}
{\small
\begin{center}
\tabcolsep=0.18cm
\begin{tabular}{cccccccccc}
\hline 
mod.  & N$_{\rm H_{Tbab}}$ & kT$_{c}$  & n         & kT$_{h}$ $^a$               & n$_{h}$         &  $\alpha$ & A$_{\rm Z}$ & F$_{0.3-10}$ & $\chi_{\nu}^{2}$/dof\\
      & $10^{22}$        & keV     & $10^{-4}$ &                       & $10^{-4}$     &           &             & $10^{-13}$       &   \\
      & cm$^{-2}$        &         &           & keV                   &               &           &             & erg/cm$^2$/s  &   \\         
\hline 
cemek & $0.05\pm0.02$    & -       &   -       & $19\pm_{5}^{10}$  & $2.4\pm0.5$ & $0.56\pm^{0.19}_{0.23}$ & $0.83\pm^{0.43}_{0.32}$ & $3.1\pm0.6$ & 1.04/94 \\
2mek  & $0.027\pm0.013$  & $0.66\pm0.04$ & $0.14\pm_{0.04}^{0.08}$ & $6.1\pm_{0.6}^{0.9}$ & $1.5\pm0.1$      & - & $1.1\pm0.4$ & $3.0\pm0.3$ &  1.03/93 \\
\hline
\end{tabular}  
\label{tab:spec}                      
\end{center}} 
\begin{flushleft}
$^a$ For cemekl, it is the maximum temperature.\\
\end{flushleft}                              
\end{table*}

\begin{table*}
\caption{Spectral parameters for the bright and faint phases, where interval B corresponds to the two stronger bright phases, b to the two weaker bright phases, and f to all faint phases (excluding the eclipse intervals). 
N$_{\rm H_{Tbab}}$ and A$_{\rm Z}$ are fixed to their average spectrum best-fitting values. The unabsorbed 0.3--10 keV are also reported. Uncertainties are at $1\sigma$ confidence level.}
{\small
\begin{center}
\tabcolsep=0.18cm
\begin{tabular}{cccccccccc}
\hline 
mod.  & N$_{\rm H_{Tbab}}$ & kT$_{c}$  & n$_{c}$         & kT$_{h}$ $^a$               & n$_{h}$         &  $\alpha$ & A$_{\rm Z}$ & F$_{0.3-10}$ & $\chi_{\nu}^{2}$/dof\\
Interval      & $10^{22}$        & keV     & $10^{-4}$ &                       & $10^{-4}$     &           &             & $10^{-13}$        &   \\
      & cm$^{-2}$        &         &           & keV                   &               &           &             & erg/cm$^2$/s &    \\         
\hline 
cemek$^b$ &&&&&&&&&\\
B  & 0.05    & -       &   -       & $27.4\pm_{4.3}^{5.4}$  & $5.2\pm2.0$ & 0.56 & 0.83 & $7.1\pm0.3$  & 1.06/55 \\
b & 0.05    & -       &   -       & $15.1\pm_{2.3}^{2.9}$  & $2.7\pm1.2$ & 0.56 & 0.83 & $3.5\pm0.2$ & 0.59/37 \\
f & 0.05    & -       &   -       & $10.3\pm_{2.5}^{3.7}$  & $0.8\pm0.1$ & 0.56 & 0.83 & $1.0\pm0.2$ & 0.98/17 \\
\hline
2mek &&&&&&&&&\\
B  & 0.027  & $0.82\pm0.10$ & $0.27\pm0.07$ & $7.1\pm_{1.2}^{1.7}$ & $3.3\pm0.2$      & - & 1.1 & $6.7\pm0.4$  & 1.08/53 \\
b  & 0.027  & $0.67\pm0.10$ & $0.19\pm0.03$ & $5.7\pm_{0.9}^{1.3}$ & $1.6\pm0.1$      & - & 1.1 & $3.3\pm0.3$ & 0.69/35 \\
f  & 0.027  & $0.63\pm0.10$ & $0.08\pm0.02$ & $5.0\pm_{1.5}^{3.7}$ & $0.5\pm0.1$      & - & 1.1 & $0.9\pm0.2$ & 0.96/15 \\
\hline
\end{tabular}  
\label{tab:spec_vs_t}                      
\end{center}} 
\begin{flushleft}
$^a$ For cemekl, it is the maximum temperature.\\
$^b$ Also $\alpha$ is fixed to the average spectrum best-fitting value.\\
\end{flushleft}                              
\end{table*}

\section{Discussion}
\label{sec:discus}
The X-ray observation of J0658 with the detection of the X-ray eclipses allowed to refine the orbital ephemeris and enabled to constrain the system parameters and to obtain information on the magnetic field geometry of the accreting WD (Table \ref{tab:binary}). 

\subsection{Binary system parameters}
\label{sub:binary}

\begin{figure}
    \centering
    \includegraphics[angle=0,width=9cm]{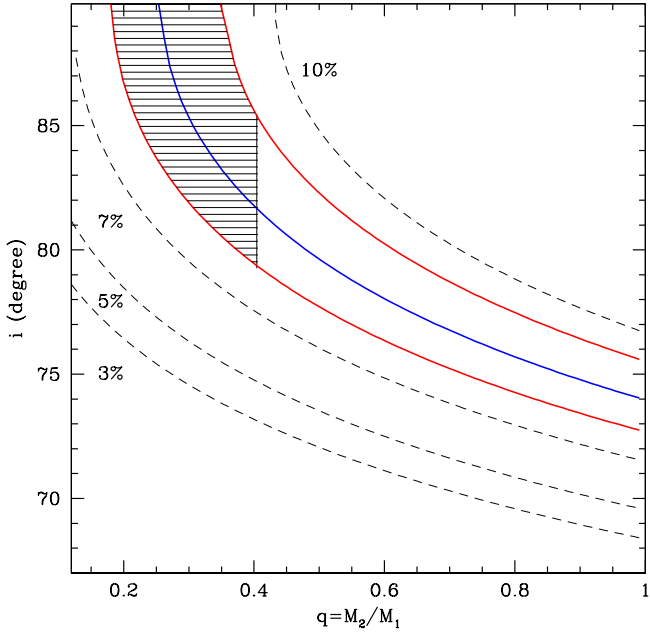}
    \caption{The eclipse fraction (its duration compared to the orbital period) as a function of the mass ratio and the orbital inclination. The blue and red lines indicate the range allowed by  the measured eclipse fraction and uncertainties  respectively. The dashed black lines indicate the inclinations versus mass ratio for different eclipse fractions. The shaded area corresponds to the limits set by the donor and primary masses (see Section \ref{sub:binary}
 for details).}
    \label{fig:eclipse_q_i}
\end{figure}

\begin{figure}
    \centering
    \includegraphics[angle=-90,width=9cm]{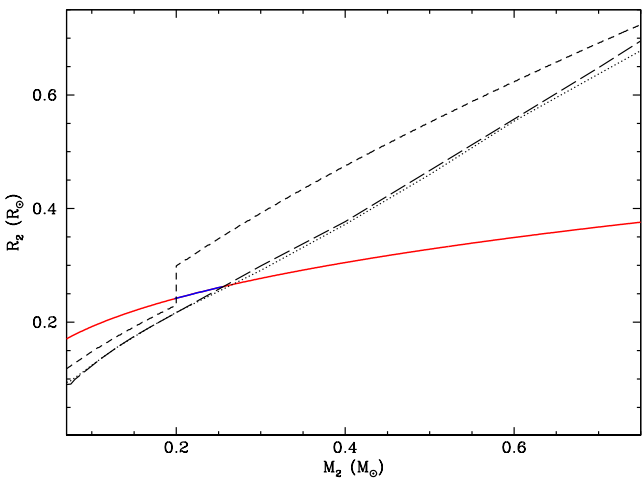}
    \caption{The M-R relation adopting a Roche-lobe filling secondary star 
and a binary orbital period of 2.38\,h (red line). The  M-R relation for 
low-mass main sequence stars  derived from the evolutionary models of 
\citep{baraffe15} at 1\.Gyr (dotted line) and  5\,Gyr (long-dashed line) 
and from the semi-empirical evolutionary sequence for CV donors by 
\citep{knigge11} (short-dashed line). The points of intersection with
the sequences give the likely range (blue line) of the  secondary star 
parameters.}
    \label{fig:mass_rad_sec}
\end{figure}

The time elapsed from the first to the 3rd contact (or from the 2nd to the 4th contact) corresponds to $\rm \Delta \phi= 0.087\pm$0.008. This allows to place constrains on the donor star radius in units of binary separation:
$ (R_2/a)^2 = (sin^2(\pi\,\Delta \phi) + cos^2(\pi\,\Delta \phi) cos^2(i)$ \citep{horne82}, where $\rm R_2$ is the donor star radius, $a$ is the binary separation, and $i$ is the binary inclination. Since J0658 is found to be accreting, the secondary star should be filling its Roche-lobe.  This equation, together with the equivalent Roche lobe radius approximation by \cite{eggleton83}, defines an unique relation between the binary inclination $i$ and the mass ratio $\rm q=M_2/M_1$. In Figure \ref{fig:eclipse_q_i} the observed eclipse length (blue line) is reported together with its uncertainties (red lines). Different eclipse fractions allow for a limited range of q-i values (black lines). 
To constrain the q-i values, the density of a secondary star filling its Roche-lobe was used: $\rm <\rho>= 3\,M_2/4\,\pi\,R_2^2 =  3\,\pi/0.459^3 \, G\, P_\Omega^2$ \citep{faulkner72}, which allows to put constrains on the radius of the secondary star for a  given mass. 
Then, to get an estimate of the donor mass and radius, the donor star mass-radius (M-R) relation adopting the 2.38\,h orbital 
period is compared with the M-R relations for late-type main sequence stars 
as derived from the evolutionary models by \cite{baraffe15} at 1\,Gyr and 
5\,Gyr and for donors in CVs as derived from the semi-empirical sequence by 
\cite{knigge11}.  These are displayed in Figure\,\ref{fig:mass_rad_sec}. 
The points where the M-R relation intersects these sequences 
define the likely ranges for the donor mass and its radius: 
$\rm M_2 = 0.2-0.25\,\rm M_{\odot}$ and $\rm R_2=R_{L2}= 
0.24-0.26\,R_{\odot}$, respectively. Adopting as a lower limit to the shock 
temperature  kT$_{\rm max}$, as derived from the X-ray spectral fits of 
interval B, using its $1\sigma$ uncertainty (23--33\,keV; 
Table\,\ref{tab:spec_vs_t}, first row), a minimum WD mass  in the range 
0.62--0.76$\rm\,M_{\odot}$ is derived. 
Then, the mass ratio is limited to $q< 0.4$, for $M_{WD}>0.6\rm\,M_{\odot}$ 
and $\rm M_2=0.25\,M_{\odot}$. In the q-i plane (Figure\,\ref{fig:eclipse_q_i}), 
the eclipse fraction limits the binary inclination in the 
range $\rm 79^o\lesssim i \lesssim 90^o$ and the mass ratio in 
the  range 0.18$<$q$<$0.40 
(shaded region). 
The latter, for a donor  mass of $0.2-0.25\,\rm M_{\odot}$, 
would give $\rm 0.5 < M_{WD}<1.4\,M_{\odot}$. 
A massive WD is not favoured, as it would imply a shock temperature 
higher than 100\,keV.
We then conservatively adopt $\rm M_{WD}$=0.6--1.0$\rm\,M_{\odot}$.
From the inferred inclination and donor radius, a binary orbital 
separation in the range 0.69--1.18 $\rm R_{\odot}$ is derived.

\subsection{The X-ray emitting region}

The X-ray modulation reveals the typical bright and faint phases seen in the 
polars, which are produced by the accretion flow above the main (or upper) 
pole that comes into (bright phase) and out of (faint phase) view if the 
magnetic and rotation axes are offset by an angle $\beta$ defined as the 
magnetic colatitude  \citep{Cropper88}. The non-zero count rate during the 
faint phase could suggest either that the accreting upper pole does not 
completely disappear behind the WD limb or that a second emitting region 
is present. The length of the faint phase and the derived inclination  
can be used to restrict the range of values of the magnetic colatitude 
of the main pole as cos\,$i$ = cos\,($\rm \pi\,\Delta\phi_{faint})\, 
tan\,\beta$, \citep{Cropper90}. The X-ray faint phase is evaluated in 
interval B, where the modulation is better defined, and is 
determined as the  
phase range where the count rate is below the average value:    
$\rm \Delta \phi_{faint}= 0.42\pm0.03$.  Using $i$ = 79-90$^{\rm o}$, 
the magnetic colatitude, neglecting any height and width 
of the column, is restricted to $\rm \beta \lesssim 50^{\rm\,o}$. Here we note
that the colatitude is undetermined for $i$=90$^{\rm o}$. $\beta$ could be 
larger if there is a vertical or horizontal extent of the emission region 
\citep[e.g.][]{Vogel08}. 

For $\beta < 90^{\rm\,o}$, the main accreting pole is on the same 
side of the orbital plane as  the line of sight of the observer.  
The azimuth $\psi$ is the angle 
from the line centres of the two stars to the main pole projected onto 
the orbital plane. This can be defined as 
$\rm \psi = \phi_{eclipse} - \phi_{max,cent}$, 
where $\rm \phi_{eclipse}$ is the phase of the eclipse centre 
and $\rm \phi_{max,cent}$ the phase of the centre of the maximum, 
i.e. when the accretion column is closest to the line-of-sight  
\citep{Cropper88}. We then derive $\psi\sim 14^{\rm\,o}$.

The decline to the eclipses (1st to 2nd contact) is found to be different in interval B and b. It ranges from 42--53\,s during interval B, while it is longer (120--162\,s) during b. The rise from the eclipses (3rd to 4th contact) is instead longer in B (2.3--3.7\,min) than in b, where is only 44\,s and 74\,s. The differences are likely due to the low S/N of the data, preventing quantitative estimates of the size of the X-ray emitting region.

The spectral analysis shows that the PSR emitting region is highly variable over time. The X-ray flux of the bright phase changes by a factor of two from cycle to cycle, whilst the minimum flux is essentially constant. During the faint phase the flux is $\sim$ 3-6 times lower than that during the bright phases (interval b and B, respectively). Although it is not possible to precisely assess the temperature structure of the PSR with present data and the adopted spectral model, the fact that lower temperatures are found during the bright phases of lower intensity (interval b) might suggest that the PSR is not stationary and could adjust itself according to the instantaneous mass accretion rate. 

\subsection{The accretion and stellar components }
\label{subs:s_comp} 
J0658 has been observed by {\it Gaia} \citep{gaia18a} that in DR2 has provided a precise parallax of  $\pi=4.75\pm0.05$ mas, accurate at $\sim 1\%$. With this accuracy the direct inversion is well justified giving D=211$\pm$2\,pc. We however also derive the distance to the source using the weak distance prior that varies as a function of Galactic longitude and latitude according to the Galactic model described in \citep{bailer-Jones18}\footnote{ http://gaia.ari.uni$-$heidelberg.de/tap.html} and find D$=209^{+3}_{-2}$ pc and a scale length L$=1573$ pc. We adopt the latter as best estimate of the distance since it is independent of assumptions on physical properties of individual stars. This close distance may explain the detection of this source in hard X-rays, similarly to the case of the other twelve polars that are also found within 500\,pc.
 
The extinction in the optical band, as derived from the hydrogen column density $\rm N_H$ of the X-ray spectral fits, is  $\rm E(B-V)$=0.08. For a donor mass M$_{2}=0.2-0.25\rm\,M_{\odot}$ and radius R$_{2}=0.24-0.26\rm\,R_{\odot}$ both the CV semi-empirical sequence \citep{knigge11} and the low-mass evolutionary models \citep{baraffe15} predict a M4 spectral type with effective temperature of $\sim$3300\,K. The expected absolute magnitude in the V-band is 12.3--12.4\,mag for both models, which in turn predicts an apparent extincted V magnitude of 19.3--19.4\,mag, fully consistent with the observed magnitude 
(19.4$\pm$0.3) during the eclipse.  The R-band magnitude during the eclipse is $\sim$18.5\,mag \citep{halpern18}.  Again, using E(B-V), the distance and predicted absolute  magnitude R=11.4--11.6\,mag, the expected apparent R-band magnitude is indeed 18.2--18.4 mag. All this confirms the donor star parameters derived above. 

\begin{figure}
    \centering
    \includegraphics[angle=0,width=8.5cm]{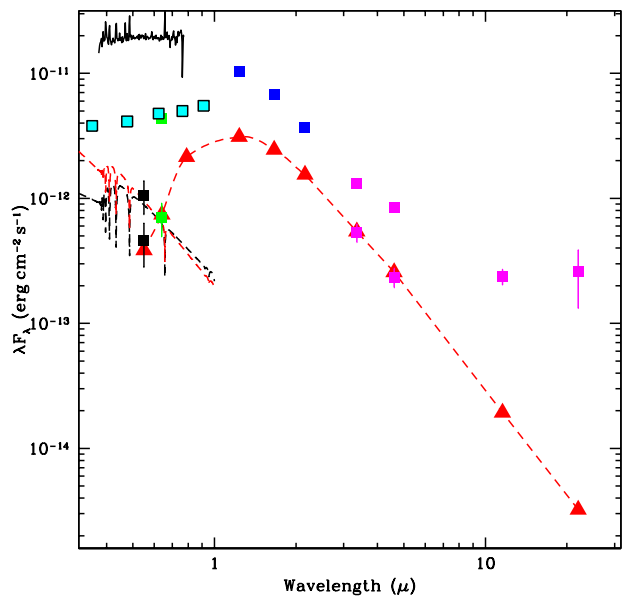}
    \caption{The multi-epoch SED of J0658 constructed using the OM  V-band photometry (black points) during and outside the eclipse, together with the R-band fluxes (green points) during and outside the eclipse \citep{halpern18} and the optical spectra \citep[black curve;][]{rojas17} or alternatively
    the SDSS (cyan squares) photometry (note the flux variability at different epochs), and the 2MASS (blue squares) and WISE photometry (magenta squares). The eclipse fluxes in the W1 and W2 bands are also reported in magenta. Overplotted as black and red dashed lines are the DA model atmospheres for a WD mass of 0.6 and 1.0\,$\rm M_{\odot}$ and temperatures of 12 kK and 22\,kK, respectively  \citep{koester10}. Red triangles are instead the predicted optical (VRI) and nIR (JHK) fluxes for a 0.2\,M$_{\odot}$ and T$_2=3300$ K donor star   \citep{baraffe15} and the predicted WISE W1, W2, W3, and W4 fluxes for a main-sequence star at the same temperature \citep{pecaut13}, together with a dashed red line to help readability. While during the eclipse the fluxes well match those predicted by the late-type companion models, the out-of eclipse fluxes, show an excess (Section \ref{subs:s_comp}).}
    \label{fig:sed}
\end{figure}

The epoch of the \xmm\ observation, due to the lower accretion rate, is more favourable to get a constrain on the direct emission from the WD atmosphere. Note that effective temperatures of WDs in polars have been found systematically lower than those in non-magnetic CVs \citep{araujo05}, with values ranging from 10\,kK to 20\,kK and for P$_{\Omega}<3$ h, up to $\sim$14\,kK.
The lack of colour photometry and/or spectroscopy allows to only derive an upper limit to the WD effective temperature. For this purpose we consider M$_{WD}$=0.6--1.0$\rm\,M_{\odot}$ and R$_{WD}=5.5-8.5\times10^{8}$\,cm, using the WD mass-radius relation of \cite{nauenberg72}. We use the DA WD model atmospheres by \cite{koester10}, adopting log\,g=8.0 and varying the effective temperatures in the range 10--24\,kK, and scale the spectral distributions (SED) using the range for the WD radius and the distance to the source reported above. We then set the upper limit  considering that the WD photospheric emission must be lower than the observed V-band flux outside the eclipse, that gives T$_{WD}<12-22$ kK (Figure \ref{fig:sed}).

\begin{figure}
    \centering
    \includegraphics[angle=0,width=8.5cm]{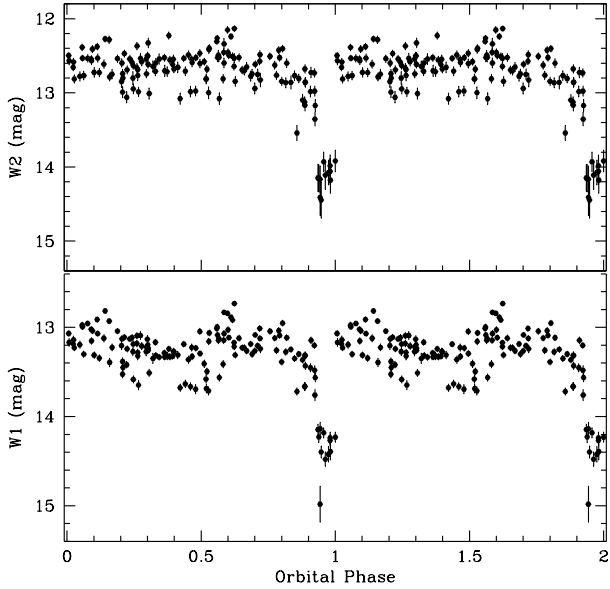}
    \caption{The NEOWISE (April 2014 - October 2018) folded light curves of J0658 in W1 (3.4$\mu$) and W2 (4.6$\mu$) bands at the 2.38\,h orbital period displaying eclipses also in the nIR. The exposure for each point is 7.7 s.}
    \label{fig:wise_lc}
\end{figure}

J0658 is also detected at  near-IR (J, H, and K) wavelengths in the 2MASS survey \citep{skrutskie06} as 2MASS\,J06580591-1744249, as well as in the four (W1, W2, W3 and W4) bands in the WISE survey \citep{cutri13} as WISE\,J065805.87-174424.5. We then built a broad-band SED from 0.35 to 22\,$\mu$ using the SDSS u, g, r, i, and z photometry and also include the optical spectrum from \cite{rojas17} (2011 April 20), the OM V and R-band measures and the 2MASS and WISE catalogue magnitudes (all measures are corrected for extinction with  E(B-V)=0.08), keeping in mind that they uncover different epochs (1998 for 2MASS, 2008 for SDSS and 2010 for WISE). The SED is displayed in Figure \ref{fig:sed}. 
Since J0658 has  additionally been observed by the reactivated WISE  
mission\footnote{WISE spacecraft was reactivated in December 2013 and named 
NEOWISE, https://irsa.ipac.caltech.edu/Missions/wise.html.} between April 3, 
2014 and October 12, 2018, the single 7.7\,s exposure data in the 
two W1(3.4\,$\mu$) and W2(4.6\,$\mu$) bands were retrieved. When folded 
at the refined ephemeris (Section \ref{subsubsec:ephem}, 
Table \ref{tab:binary}), the NEOWISE photometry clearly shows 
eclipses at these wavelengths (Figure\,\ref{fig:wise_lc}), where the 
drop in magnitude is $\rm \Delta W1 \sim$ 1.0\,mag
and $\rm \Delta W2\sim$ 1.2\,mag. We are unable to further improve
the orbital period with these data since the eclipse is covered only in a few
epochs with a handful of exposures.
The out-of-eclipse phases are characterised 
by a large variability, likely due to the multi-epoch coverage of the 
observations and thus not used in the SED. The eclipse measures in the 
V, R and W1, W2 bands are also shown in Figure \ref{fig:sed}. 
A comparison with the predicted broad-band fluxes from a 
0.2\,$\rm M_{\odot}$ donor at 3000\,K from \citep{baraffe15} clearly 
shows the agreement with all the eclipse fluxes. Instead, the fluxes 
out-of-eclipse show an excess extending down to 22\,$\mu$.  The excess 
at optical wavelengths could be a mixture of the accretion stream and 
cyclotron contributions, while the latter could be dominant in the nIR. 
Indeed, \cite{harrison15} showed that those polars with strong nIR excess (down to $\sim 12-22\,\mu$) are dominated by the fundamental and/or lower harmonics. In particular those systems having the fundamental in the longer-wavelength WISE band (W3) should be low-field ($\rm B<10\,MG)$ polars. We note that J0658 could be a low-field polar.

To evaluate the excess of flux, we subtracted from the observed out-of-eclipse fluxes in each band the contribution of the WD or donor star. More specifically for the Sloan u and g filters we subtracted the model atmosphere flux of a WD at 22\,kK and of 1.0\,$\rm M_{\odot}$ here assumed as an upper limit, while for the longer wavelength bands we subtracted the predicted fluxes for a donor of 0.2$\rm M_{\odot}$ and 3300\,K. Here we do not include the V-band data of September 2018, since J0658 was in a faint state. This gives
an integrated flux of $\rm \sim 1.2 \times 10^{-11}\,erg\,cm^{-2}\,s^{-1}$. If the optical spectrum from \cite{rojas17} is used, instead of the Sloan data, an integrated flux of $\rm \sim 2.2 \times 10^{-11}\,erg\,cm^{-2}\,s^{-1}$ is obtained. The corresponding luminosity is $\rm \sim1.2\times 10^{32}\, erg\,s^{-1}$.
The latter could be regarded as an estimate of the high state accretion luminosity at optical/IR wavelengths. Comparison with the X-ray luminosity is then performed using the higher state \Swift \, observations in 2009 and 2013 for which the derived flux (from 0.01 to 10\,keV) is  $\rm \sim7.8\times 10^{-11}\,erg\,cm^{-2}\,s^{-1}$ obtained from the simple power-law fit (Sect.\,3.2), and the average BAT spectrum for which the derived (10--200 kev) flux using a simple power-law model ($\Gamma=2.0$) is $\rm \sim 1.1 \times 10^{-11}\,erg\,cm^{-2}\,s^{-1}$. The corresponding 0.01-200 keV X-ray luminosity is then:  $\rm \sim4.7\times 10^{32}\, erg\,s^{-1}$, which is a factor of $\sim$4 larger than that derived in the optical/IR range. We then estimate the mass accretion rate assuming ${\rm L_{accr} = L_{x} + L_{opt} = G M_{WD}} \dot{\rm M}/\rm R_{WD}$. 
For $\rm M_{WD}\sim 0.6-1.0\,M_{\odot}$, the total accretion luminosity of $\rm \sim5.9\times 10^{32}\, erg\,s^{-1}$ gives  
$\dot{\rm M} \rm \sim3.8-9.6\times10^{-11}\,M_{\odot}\,yr^{-1}$. 
This value is lower than those of CVs above the 2--3\,h orbital period gap and rather consistent with those of systems below it \citep{knigge11}. Considering that J0658 is located in the middle of the  gap, the mass transfer rate is expected to drastically drop when entering the gap. 
CVs can be born inside the gap and contribute to the number of systems observed there \citep{goliasch_nelson15}. Another possibility is that J0658 is one of the polars
that never had an effective magnetic braking,
and hence it evolved, without a significant break, from a longer initial period \citep{Webbink02}.

J0658 is then the second hard X-ray discovered polar in the orbital period gap, together with SWIFT\,J2218.4+1925 (2.16\,h) \citep{bernardini14}. So far the total number of polars in the gap amounts to 34 systems \citep[][updated 2016 catalog]{ritter_kolb}, $\sim 26\%$ of the whole polar population. The lack of a well-defined gap in mCVs \citep{Webbink02} was ascribed to the the tight coupling between the donor and WD magnetic fields, reducing the magnetic braking as the wind from the secondary star may be trapped within the WD magnetosphere.

\begin{table}
\caption{Binary system parameters for J0658.}
\begin{center}
\begin{tabular}{lc}
\hline 
Orbital period & 0.09913398(4)\,d \\ 
Eclipse egress (BJD) & 2458381.333958(65)  \\
\hline
Orbital inclination ($i$) & 79--90$^{\rm\,o}$   \\   
Mass Ratio (q)  & 0.18--0.40  \\
WD Mass$\rm ^a$ ($\rm M_{WD}$) &  $>$ 0.6\,$\rm\,M_{\odot}$  \\
Secondary Mass (M$_{2}$) & 0.2--0.25$\rm\,M_{\odot}$  \\
Secondary Radius (R$_{2}$) & 0.24--0.26$\rm\,R_{\odot}$ \\
Secondary Spectral Type & M4 \\
Distance (D) & $209^{+3}_{-2}\rm\,pc$ \\
Main Pole colatitude ($\beta$) & $\lesssim 50^{\rm\,o}$  \\
Main Pole azimuth ($\psi$) & $\sim14^{\rm\,o}$ \\ 
\hline              
\end{tabular}
\label{tab:binary}
\end{center}
\begin{flushleft}
$\rm ^a$ This represents a lower limit to the real WD mass.
\end{flushleft}
\end{table}  

\section{Conclusions}
\label{sec:conc}

J0658 is the thirteen hard X-ray selected polar discovered so far. The 
presence of eclipses with a period of 2.38\,h locates it in the middle 
of the 2--3\,h orbital period CV gap. Its X-ray emission is modulated at 
the orbital period and the intensity of the modulation is variable (factor 
of two) from cycle to cycle, a signature of a non-stationary mass accretion 
rate. The X-ray luminosity it is also found to be highly variable on long 
(yrs) timescales (factor of sixty) and \xmm\ caught it at the lowest state 
ever observed. The X-ray spectrum is thermal and consistent with a 
multi-temperature structure, as observed in many magnetic systems. 

The eclipses also allow to constrain the binary inclination between 79$^o$ 
and 90$^o$, the mass ratio $\rm q=0.18-0.4$ and orbital separation 
$\rm a=0.69-1.18\,R_{\odot}$.  From {\it Gaia} parallax the distance is 
firmly set at $209\pm^{3}_{2}$\,pc. We estimate a donor star mass of 
0.2-0.25$\rm M_{\odot}$ assumed to fill its Roche-lobe at a temperature 
of 3000\,K and a lower limit to the WD mass of 0.6$\rm M_{\odot}$, with 
an upper limit to its effective temperature of 12--22\,kK.  The presence 
of a non-negligible optical/IR excess, suggests a contribution of cyclotron 
radiation at these wavelengths. J0658 may host a weakly magnetic WD 
($\rm B\lesssim10$\,MG). 
We tentatively estimate a magnetic field orientation of the 
main accreting pole with colatitude $\beta\lesssim 50^o$  
and an azimuth of $\sim14^o$.  Spectropolarimetry will be crucial to 
determine the magnetic field strength and its topology for this polar. 
With thirteen systems found as hard X-ray polars, the possibility that this sub-class of mCV could be a non-negligible contributor to the hard X-ray low-luminosity Galactic  population ($\rm \lesssim 10^{32}\,erg\,s^{-1}$) should be further investigated.

\section*{Acknowledgements}

FB is founded by the European Union's Horizon 2020 research and innovation programme under the Marie Sklodowska-Curie grant agreement n. 664931. DdM and NM acknowledge financial support from the Italian Space Agency and National Institute for Astrophysics, ASI/INAF, under agreements ASI-INAF I/037/12/0 and ASI-INAF n.2017-14-H.0 and from INAF SKA/CTA Presidential Decree N. 70/2016.  This work is based on observations obtained with \xmm, an ESA science mission with instruments and contributions directly funded by ESA Member States, with \Swift, a NASA science mission with Italian participation, and with \textit{Gaia}, a ESA mission. \textit{Gaia} data are processed by the Data Processing and Analysis Consortium (DPAC). We thank the anonymous referee for the useful comments. We acknowledge useful discussion with Carlo Ferrigno. 

\bibliographystyle{mnras}
\bibliography{biblio} 

\bsp	
\label{lastpage}
\end{document}